# Reducing Risk and Accelerating Delivery of a Neutron Source for Fusion Materials Research

E. Surrey[*a], M. Porton[a], T. Davenne[b], D. Findlay[b], A. Letchford[b], J. Thomason[b], S.G. Roberts[c], J. Marrow[c], A. Seryi[c], B. Connolly[d], H. Owen[e]

[a]EURATOM/CCFE, Abingdon, OX14 3DB, UK. [b]STFC Rutherford Appleton Laboratory, Harwell, OX11 0QX, UK. [c]University of Oxford, Oxford, OX1 3DP, [d]UK, University of Birmingham, Edgbaston, B15 2TT, UK, [e]University of Manchester, Manchester, M13 9PL, UK

*Abstract—* The materials engineering data base relevant to fusion irradiation is poorly populated and it has long been recognized that a fusion spectrum neutron source will be required, the facility IFMIF being the present proposal. Re-evaluation of the regulatory approach for the EU proposed DEMO device shows that the purpose of the source can be changed from lifetime equivalent irradiation exposure to data generation at lower levels of exposure by adopting a defence in depth strategy and regular component surveillance. This reduces the specification of the source with respect to IFMIF allowing lower risk technology solutions to be considered. A description of such a source, the Facility for Fusion Neutron Irradiation Research, FAFNIR, is presented here along with project timescales and costs.

## 1 Introduction

The need to establish a facility capable of irradiating materials with a neutron spectrum that mimics that generated by a fusion power plant was identified in the 1980s. The historical role advocated for a fusion relevant neutron source includes population of the materials database with engineering relevant information, provision of 14MeV neutron irradiation data to validate and calibrate alternative irradiation techniques and qualification of materials to a lifetime use equivalent of 150dpa. The International Fusion Materials Irradiation Facility (IFMIF) [1,2] is the result of an assessment of different concepts intended to provide this. The resulting IFMIF specification requires machine availability of 70% from two accelerators operating at the highest cw power recorded, imposing 1GWm$^{-2}$ of beam power on a flowing lithium target. These demands are challenging and present a high technological risk. Although funded by the European and Japanese ITER members under the Broader Approach, no timetable is foreseen that will deliver materials testing data on a timescale commensurate with the start of DEMO construction proposed in the newly-adopted EU Fusion Roadmap [3].

This obviously impacts upon the design program for power plants and has prompted this study to assess, within the context of regulatory licensing and the engineering materials perspective, the actual requirements for the neutron source to precede this Roadmap DEMO milestone. This approach shows that a facility of reduced intensity, based on (near-) available technology, can provide a valuable resource if realized on a suitable timescale.

## 2 Requirements of a neutron source

The role of a neutron source within the fusion program is primarily to populate the materials database with engineering design relevant information. Within this role is the provision of 14MeV irradiation data to validate and calibrate the more readily available fission and ion irradiation data and to strengthen predictive modeling capability. The original intention of IFMIF was the qualification of candidate materials up to a full lifetime use (assumed in [1] to be 20 years), equivalent to approximately 150dpa [1, 2]. This need is based on the perception that such qualification is necessary for regulatory licensing of a fusion power plant. A re-assessment of the regulatory requirements and those of the engineering materials indicates that some of these original specifications can be relaxed

*2.1 Requirements Determined by Regulatory Considerations*

The regulator will insist that materials used for the construction of the radiological confinement boundary are demonstrably safe over the lifetime of the plant whilst investors and stakeholders will seek reassurance of the integrity of the whole plant.

Application of "defense in depth" strategies, as adopted by ITER [4] should allow the regulatory requirements to be met without the need for a prolonged irradiation qualification campaign. By defining the primary confinement boundary to be the vacuum vessel and its extensions, the in-vessel components such as the plasma facing first wall, tritium breeder blankets and divertor are no longer part of the radiological control. This circumvents the need of end-of-life testing for the in-vessel components and the inherent difficulty in achieving high dpa material, joint and component irradiation by a 14MeV neutron source.

This approach necessitates that the vacuum vessel material must be adequately characterized (along with materials comprising any of its extensions such as auxiliary heating systems). Simulation shows the high energy neutron flux at the vacuum vessel wall is over 10$^4$ lower than at the first wall and considerably softer with less than 30% of the flux having

---

energies above 0.1MeV. The flux below this energy is reduced by ~600 compared to the first wall [5] so that over the 30 year lifetime specified for the DEMO device the exposure to the main vessel will be of the order 0.2dpa. Qualification of materials to this exposure would not require lengthy irradiation times in even a modest flux source. Further mitigation can be provided by additional confinement structures so the requirements to be met by the neutron source become primarily the provision of data to assure investment protection and engineering design substantiation.

This will be difficult to achieve in the absence of many years irradiation by a 14MeV neutron source. In addition, the proving of joining techniques and component assemblies will be severely limited in an accelerator driven source due to volumetric constraints. This inherent uncertainty in the material properties under irradiation draws many parallels with the $20^{th}$ century realization of first-generation fission plants, particularly in the realm of design criteria and their interaction with safety and materials activities.

Given the substantial gaps in understanding materials performance within fission reactors and the absence of nuclear design codes, a pragmatic approach was taken to facilitate the design and continued operation of the plants, re-assuring the regulator and enabling the long-term development of fission design criteria. Most importantly, the safety case was formulated with key statements to ensure continued plant operations were dependent upon resistance to failure but that neither advance knowledge of end-of-life material performance nor exhaustive experiences of the failure modes were required.

Formulating the safety case in this way meant that rather than exhaustive testing and development programs in advance of the build, ongoing demonstration of regulatory compliance was instead dependent on continuous in-service assessment to demonstrate an acceptably low probability of failure to the regulator. This was achieved by extensive surveillance schemes; withdrawal of material and joint specimens at periodic intervals allowed tracking of changes in properties and development of models to allow interpolation and extrapolation with confidence. Understanding of effects of each variable (and physical processes) over the life of the device facilitated good predictions and ultimately regulator confidence. This multi-faceted approach including safety expertise, dedicated experiments and supporting materials modeling, allowed the licensing of first-of-a-kind plant types. Substantial improvements in understanding of both material behavior and mechanisms of failure over the life of the project were ultimately iterated into the development of new design criteria to guide the design of upgrades and future plants.

This early fission experience provides a number of important lessons for fusion:

(i) the approach to licensing cascades into the safety case and important decisions on the scope of design criteria development in advance of the plant build

(ii) design criteria and their development must be undertaken in close co-operation with dedicated supporting materials experiments and materials modeling activities

(iii) complete understanding of the environment is not needed; therefore end of life fusion neutron irradiation is not required before the design and build of DEMO. Instead only an insight into the effects is needed with margin provided for the inevitable 'unknown unknowns' that will be revealed during the lifetime of the project

(iv) accelerated testing programs pursued in parallel to operations are important to facilitate long-term learning.

To minimize the scope of work required to facilitate the realization of DEMO, this pragmatic approach, adjusted for a modern context and regulatory system, offers many attractions. However, minimizing the amount of work in advance does raise technical risks for the design. In particular, designs may be susceptible to crippling 'unknown unknowns' such as new failure modes and their interactions, which could serve to reduce component lifetime and therefore plant availability.

*2.2 Requirements Determined by DEMO Operation*

The purpose of the EU DEMO device was re-assessed in 2012 [6], emerging as a technology demonstrator capable of delivering 500MWe but with limited availability of 30%. Furthermore, it is envisaged that the plasma facing first wall (~2MWm$^{-2}$ neutron flux) components will be replaced after an exposure equivalent to 20dpa in steel, a calendar time equivalent to approximately 4 years assuming a damage rate of 15dpa per full power year (fpy) and 30% availability. This relaxes the operating characteristics of the neutron source significantly from the IFMIF requirement.

*2.3 Requirements Determined by Materials Database*

A re-assessment of the neutron source requirements from an engineering materials perspective shows that some of the original requirements can be relaxed. For example, materials degradation phenomena such as irradiation creep, volumetric swelling, and phase instabilities approach saturation at damage levels above 10dpa [7] a level that can be achieved in a reasonable time frame with a less intense source of 5dpa/fpy. The onset of material embrittlement due to the transmutation production of helium and hydrogen is more difficult to quantify as this is only manifest through indirect evidence such as changes in tensile properties and the ductile to brittle transition temperature (DBTT). Experimental evidence suggests that helium concentrations of 400appm at 15dpa have no effect on the tensile properties at temperatures from 250 to 400$^0$C whilst the same concentration would increase the DBTT by some 200$^0$C as measured by a Charpy test [8]. Despite this apparent problem to detect embrittlement below 15dpa, valuable early elimination of unsuitable materials can be achieved to minimize risk to the engineering design and development programme.

Recent advances in the use of small (millimetre scale) mechanical test specimens, where the "dead areas" can be multiply-purposed for hardness testing, thermal and electrical testing, and for production of microscopy and ultra-small scale mechanical tests, allows the high flux irradiation volume (>20dpa/fpy) to be reduced from the 500cm$^3$ in IFMIF without loss of statistical value in test results.

*2.4 Purpose and Specification for the Neutron Source*

Combining the requirements of sections 2.1, 2.2 and 2.3, the purpose of the 14MeV neutron source can be defined, in priority order, as:

(i) demonstrate lifetime integrity of confinement boundary materials under relevant neutron spectrum and exposure

(ii) identify new phenomena associated with 14MeV neutron irradiation that may impact on the safety case, necessitating further investigation

(iii) provide significant contributions to the population of the engineering materials database and eliminate unsuitable candidate materials

(iv) validate and calibrate fission and ion irradiation techniques and advance the materials modeling capability for fusion without compromising the validity of the neutron spectrum.

(v) provide assurance for protection of investment for stakeholders through development of design codes

From these perspectives, the specification of the 14MeV neutron source can be summarized as:

(i) the neutron energy spectrum should mimic that expected at the "first wall" (i.e. the plasma facing inner surface of the tokamak vessel) in terms of primary recoil spectrum (PKA), He and H generation and important transmutation products and with respect to time signature.

(ii) to be applicable to the EU fusion roadmap, the facility should provide data at exposures of 30dpa before the end of 2026, implying exposure rates of approximately 10dpa/fpy to allow a number of samples to be tested to inform the engineering design phase of DEMO starting in 2021. This dictates mature or near term technology, requiring minimal development to reduce costs and expedite construction and commissioning.

(iii) to achieve availability commensurate with the timescale the facility should avoid remote maintenance wherever possible. Clearly the target and irradiation volume will require remote handling but the accelerator should be designed for hands-on maintenance, implying a low beam loss.

(v) exploitation of the irradiation volume must be optimized to ensure the potential to populate the materials database is maximized. This implies beam scanning or spreading, the use of smaller specimen sizes than proposed for IFMIF and the production of multiple test samples from each specimen.

*2.5 Assessment of Neutron Sources*

There are basically four neutron source options: spallation, stripping, beam-plasma and fission reactor and examples of the normalized neutron energy spectra from examples of each of these are shown in Fig 1, together with that from the fusion plasma. Of these only the beam-plasma

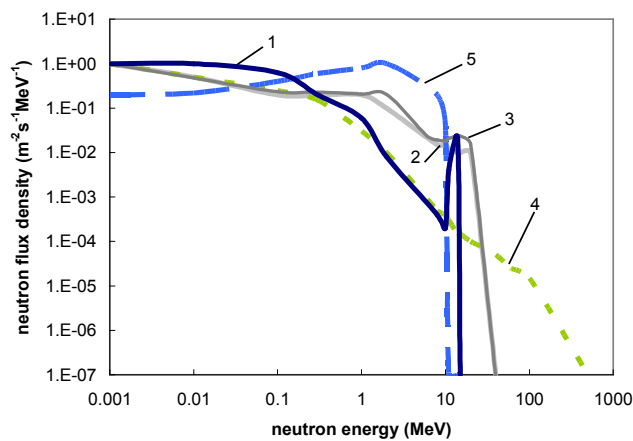

Fig. 1. Normalized neutron energy spectrum of the sources considered: 1 DEMO first wall and D-T beam-plasma , 2 IFMIF High Flux Module, 3 FAFNIR, 4 spallation source, 5 fission (PWR)

source can be constructed to provide an accurate representation of the fusion neutron energy spectrum by using deuterium and tritium in the beam and/or plasma. However this technology is relatively immature and has high capital costs due to the tritium content.

The fission reactor option provides high damage accrual rate but low He and H concentrations due to the lack of neutrons with energies above 10MeV. The spallation source is a mature technology providing high damage accrual rate but excessive He production due to the high energy content of the neutron spectrum and the co-incident protons. As a secondary consideration, spallation sources use high energy ion beams (100-1000MeV) and are inherently of short pulse length with frequencies around 10-100Hz and there are some concerns regarding the validity of applying a pulsed source to material irradiation studies for an essentially cw technology when typical timescales for microstructure evolution of irradiation damage range from $10^3$s to $10^{-6}$s and are temperature dependent.

Stripping sources, of which IFMIF is an example, use lower energy ion beams (20-50MeV) incident on a target whereby the choice of beam ion and target material dictate the neutron energy spectrum. These can produce a good simulation of the fusion neutron spectrum and provide cw irradiation but a high damage accrual rate is challenging as this implies a relatively high beam current. Of the various options available - beam into gas, liquid metal target, solid target, fluidized powder target – only the deuteron beam onto a rotating carbon target fulfils the criteria of low technical risk and readily available (or near term) technology

## 3 FAFNIR

The proposed FAcility for Fusion Neutron Irradiation Research (FAFNIR) is a stripping source based on the C(d,n) reaction utilizing a 40MeV $D^+$ beam impinging upon a rotating solid carbon target. The reaction and beam energy ensure a neutron spectrum peaked at 14MeV (Fig 1) whilst providing improved yield over protons. The choice of beam and target elements was made after an exhaustive review of neutron sources to establish the optimum neutron spectrum and technological maturity. Realizing a 14MeV neutron source on a timescale relevant to DEMO precludes a significant R&D program, so risk reduction was a major consideration in selection of the technology.

In seeking to maximize the available irradiation volume a design option of 30mA beam current was initially adopted. However this would represent a power load of 1.2MW on the target, somewhat in advance of values presently realized on existing carbon-based systems. For this reason a phased program of enhancements is proposed, summarized in Table 1 and compared with the IFMIF specification. To achieve this, the accelerator must be designed to operate at 30mA, so that only the target requires development or replacement.

It is foreseen that such a program will allow early data generation whilst developing the target in parallel. Indeed if the target development program were to begin at the conceptual design stage, it may be feasible to enter operation directly at the Near-term Baseline option.

The individual key components of the facility – ion source and accelerator, target and irradiation volume – are described in the following sections.

Table 1 Proposed Phases of FAFNIR

|  | IFMIF | FAFNIR | | |
|---|---|---|---|---|
|  |  | *Target Technology Level* | | |
|  |  | *Existing (Default)* | *Near-term (Baseline)* | *Prospective (Upgrade)* |
| *Beam* | 40MeV, 250mA | 40MeV, 2.5mA | 40MeV, 5mA | 40MeV, 30mA |
| *Target* | 10MW Liquid Li | 100kW Solid rotating C Single slice | 200kW Solid rotating C Single/multi- slice | 1.2MW Solid rotating C Various options |
| *Typical dpa/fpy in volume* | ≥1 in 6000cm$^3$ ≥20 in 500cm$^3$ ≥50 in 100cm$^3$ | ≥0.6 in 100cm$^3$ ≥1 in 50cm$^3$ ≥3.8 in 10cm$^3$ | ≥1 in 150cm$^3$ ≥1.5 in 100cm$^3$ ≥4 in 25cm$^3$ | ≥5 in 150cm$^3$ ≥7 in 100cm$^3$ ≥20 in 25cm$^3$ |

*3.1 Accelerator*

The FAFNIR accelerator configuration will be a $D^+$ ion source with dc injector to ~90keV beam energy, a radio frequency quadrupole accelerator (RFQ) to ~3MeV and a drift tube linear accelerator (DTL) continuing to 40MeV. At the proposed current of ~30mA cw the beam power would be ~1.2MW. Beam powers of ~1MW are substantial in accelerator applications; similar powers have been reached in existing facilities but at a much higher energy (and hence lower current). It is expected, from experience, that an availability (including maintenance periods) of 70% will be achievable after operational shakedown.

To avoid the need for prohibitively expensive remote handling facilities it is essential to minimize the induction of radioactivity in accelerator structures and beam lines from beam losses. To achieve similar handling strategies to the ISIS beam line [9], losses will need to be 100 times smaller. To achieve these levels of loss requires a well-controlled and low-emittance beam in a generous aperture.

The obvious technology choice for a highly reliable $D^+$ source is the off-resonance electron cyclotron resonance (ECR) source, such as developed for IFMIF [10], which has demonstrated currents over 100mA cw with low emittance, good reliability and lifetime. Reducing the accelerating gradient of the RFQ and the linac results in a longer component but offers reduced heat dissipation, making structural stability easier to maintain.

From a beam dynamics point of view, the DTL does not present any special challenges but controlling the beam losses and the power dissipation in the structure are non-trivial. A room-temperature DTL is probably the most conservative choice for the energy range ~3–40MeV although it has a significantly reduced aperture compared with superconducting options.

In order to generate a uniform (<10%) beam on target with a square profile ~60×60mm$^2$ two possible options are raster scanning, a technique widely employed at medical facilities and manipulation of the transverse beam distribution by non-linear magnetic lenses, having the advantage of delivering roughly constant power density.

Although activation of the accelerator structure will be minimized, substantial thicknesses of shielding around the accelerator will be required, of the order of 2.5m of concrete, set by the need to provide protection under abnormal conditions.

*3.2 The Target*

For a 40MeV D$^+$ beam incident on a carbon target, the beam energy is deposited in the target within 5mm of the surface. Dissipating the resultant high power densities in a stationary target would be extremely difficult but a rotating target significantly reduces the average power density and provides increased surface area for heat transfer. Calculations using TRIM suggest that for a 30cm diameter wheel with 6cm beam width a 5mA deuteron beam current would produce ~2dpa in the carbon target per six months of operation, implying that target replacement frequency can be combined with regular maintenance schedules.

FAFNIR would seek to apply the considerable worldwide experience of operating rotating solid targets to deliver a reliable initial target having precedent. By collaborating with current international development programs, the target power handling could then be enhanced both in the near- and long-term to achieve stepped upgrades in capability.

To date, single slice rotating carbon targets have been widely deployed with deposited powers of 100kW, directly applicable to the FAFNIR default specification. In the near term parallel prototyping activities in France [11] and USA [12] will prospectively provide targets capable of handling 200kW of deposited beam power. Beyond this, a dedicated target development would be required, undertaken in parallel to FAFNIR operation with the intention of upgrading the target power capability still further and achieving dpa rates compatible with end-of-life DEMO irradiations. Several target approaches should be assessed including actively-cooled multi-slice and fluidized powder targets..

*3.3 The Radiation Volume & Materials Testing*

For the upgraded FAFNIR specification of a 40MeV 30mA 30mm radius beam of uniform composition onto a carbon target, an indicative dose distribution at the mid-plane of the irradiation zone is shown in Figure 2. The volume anticipated from the preliminary FAFNIR design is between 40x40x60mm$^3$ and 50x50x60mm$^3$. Within this volume neutron dose and dose rate will vary strongly with depth; but with some scanning and beam spreading, it is expected that variation in-plane can be minimized. Irradiation temperature will also vary to some extent with position but it is expected that this can be minimized by use of differential external heating and with careful choice of location of test specimens. However, dose variation with depth will be strong over the first few millimetres and significant thereafter. This has implications for specimen types that can be used, and the packing arrangements necessary.

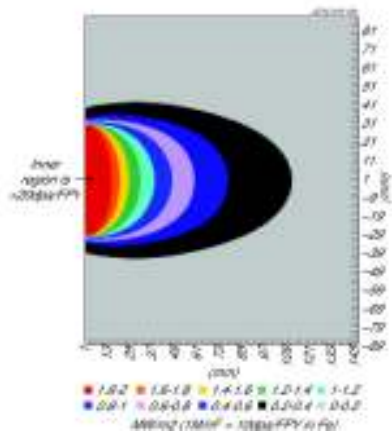

Fig 2 Estimated neutron flux in the irradiation volume of FAFNIR for 30mA beam option.

Several material types will need to be included per operational campaign and to provide statistically significant data smaller specimen types than those proposed for IFMIF will be used wherever possible. Some examples are:

(i) specimens 12x1x1mm$^3$ for pre-cracked bend samples [13]. Each iso-dose layer (~1mm) of the target zone could then contain up to 250 specimens, depending on irradiated area, providing toughness-temperature curves for up to 8 brittle materials. For more ductile metals, small punch fracture tests (8 mm diameter and 0.5 mm thickness) developed in the nuclear fission area would be more appropriate.

(ii) for metal tensile strength and fatigue testing specimens 12 x 1 x 0.4 mm$^3$ would be sufficient; the minimum dimension being still considerably larger than the grain size. Stacked "edge-on" to the neutron beam an iso-dose layer would hold up to 600 specimens providing 20 datasets.

(iii) for creep tests, sub-sized dog-bone tensile specimens of length 10mm, diameter 4.7 mm (the gage section of length 4 mm and 1.3 mm diameter), or "small punch" tests are anticipated for post-irradiation testing of creep resistance.

(iii) "dead zones" in "large" (millimetric) fracture or tensile specimens can be used for (nano) indentation testing, lift-out for transmission electron microscopy and atom-probe tomography specimen preparation, micromechanical test element production [14] and measurement of thermal conductivity, electrical conductivity and swelling.

(iv) micromechanical test methods (typical dimensions from 0.5 x 0.5 x 8μm to 10 x 10 x 100μm) will be used, subject to satisfactory validation, to supplement test data from the larger specimen types. Such tests can be carried out over a wide temperature range for derivation of yield, work-hardening and fracture data in brittle materials. This is a very rapidly developing area of study via experiment and modeling and would be exploited should the issues of transferring data from very small test scales to the engineering scale be resolved.

The large-microstructure non-metallic materials are more problematic for tensile and flexural tests; graphite and ceramic matrix composite microstructures will certainly require larger specimens than metallic materials. A suitable cross-section of specimen is likely to be approximately 3x1 mm, and length 12-15 mm, thus an iso-dose layer could only accommodate 30-50 specimens. This would allow only a sparse population of tests matrix at each dose per dedicated campaign, especially for anisotropic materials. Significant reliance will need to be placed on modeling (validated against a wide range of tests on un-irradiated materials) to make best use of these data. Nevertheless, well targeted studies are feasible within the available material volume. For non-metallic materials that would be deliberately situated in low dose regions of the DEMO device, such as windows and insulators, a larger volume of dose rate <0.2dpa/fpy is available outside the main irradiation volume, as shown in Fig 2, so significant numbers of these samples can be accommodated.

*3.4 Project Risks*

A risk assessment of the key components identified the main risk to be the inability of the rotating multi-sliced target to withstand the deposited beam power density either through temperature or stress considerations. This could be mitigated through a number of actions ranging from a development program or through manipulation of the beam. Obviously the latter option would have consequences for the exposure accrual rate. A second, lesser risk associated with the target is limited target lifetime due to in-service effects (e.g. dimensional change, build-up of localized stresses, degradation of thermal conductivity, and evaporation). These can be mitigated by informed choice of material based on experimental experience, improved modeling and development of target layout and replacement scheme to facilitate rapid replacement and to minimize down-time.

Several lesser risks were identified for the accelerator: D$^+$ source stability, cw operation of the DTL and the RF drivers and production of the uniform beam footprint on the target. Most of these could be mitigated by the early inclusion of test rig facilities in the program. In the case of the beam footprint, beam dynamics simulations with realistic particle distributions are required.

For the material data accumulation the main risks were identified as: insufficient population of the fracture test matrix for ductile metals and non-metallic composites from the use of millimetric to centrimetric-scale compact-tension specimens. In this case the experience from fission materials irradiation and from modeling will allow the use of other properties, e.g. tensile strength changes and hardness changes, microscopy of grain boundary changes, to interpolate between and extrapolate beyond data obtained.

*3.5 Project Timescale and Costs*

The project timescale for the base-line option of Table 1 is shown in Fig 3 and it is clear that to comply with the present EU roadmap for DEMO, construction would have to start soon if material irradiation to over 10dpa is to be achieved before commencement of the DEMO construction in 2030.

It is recognized that existing solid target technology is not adequate for the full performance specification of the facility and that development in this area would be needed. An aggressive target development program is therefore proposed that would seek to cooperate with other groups in order to realize a 200kW target and 5mA beam current from the outset of FAFNIR operations. Continuation of this program in parallel to the FAFNIR build, commissioning and early operational activities would enable further rapid upgrade of the facility to achieve the full envisaged capability of 30mA beam current, facilitating irradiations of ~100cm$^3$ of material in excess of 20dpa by 2030. The project plan shown in Fig 3 is considered to be realistic in terms of development, installation and commissioning timescales but will not now meet the EU Roadmap requirement of 30dpa by the end of 2026 but delivers 30dpa in early 2008 (note that the FAFNIR concept originated in 2012 at which point the 30dpa in 25cm$^3$ volume would have been achieved at the end of 2025).

Importantly the plan in Fig 3 allows time for the post irradiation analysis of the phase 1 operation (6dpa in a 25cm$^3$ volume) that will deliver early information regarding new phenomena and indicative irradiation behavior by 2023. The reference performances for FAFNIR, based on the plan in Fig 3 are given in Table 2 for a facility availability of 70%. Various options could be explored to accelerate the program, such as starting the design of target 2 in tandem with target 1 or attempting entry level at 1.2MW but the latter certainly is contrary to the philosophy of the concept as lowest risk with maximum probability of delivering results. Given that the only irradiation data obtained under fusion relevant conditions date from the RTNS campaigns in the 1980's [15] with exposure up to ~0.02dpa, the need to generate data at higher exposure levels, at the earliest opportunity, to investigate new phenomena and the possible validation of spallation, fission and ion irradiations is paramount.

Table 2 Reference performance values for FAFNIR

|  | dpa in irradiation volume (70% availability) | | |
| --- | --- | --- | --- |
| Year end | 25cm$^3$ | 100cm$^3$ | 150cm$^3$ |
| 2023 | 6 | 2.5 | 1.5 |
| 2026 | 14 | 5 | 3.5 |
| 2028 | 42 | 14.5 | 10.5 |
| 2030 | 70 | 24.5 | 17.5 |

The project costs were obtained from the reasonably detailed project breakdown structure for IFMIF given in the CDR [1]. There are several differences between the two projects, for example the IFMIF costing assumed post irradiation examination facilities would be provided by the host so were not included in the final CDR cost. These were re-introduced to the FAFNIR costing using the un-scaled figures from the initial IFMIF CDR specification. Items such as the lithium purifying plant to remove tritium and beryllium and the positive pressure argon atmosphere for the lithium target loop buildings are not necessary for FAFNR, so were excluded from the costing (although some tritium will be produced from D-D reactions).

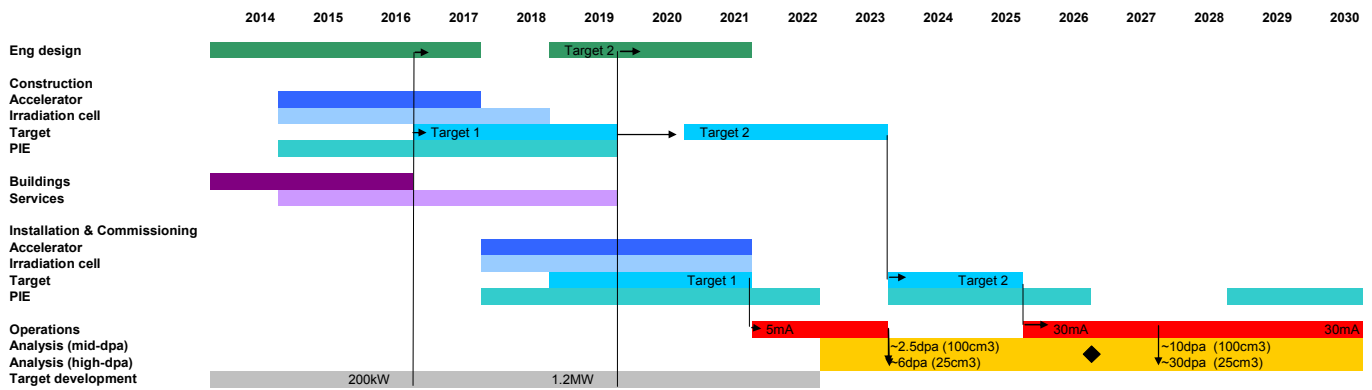

Fig 3 Timescale for the base-line 200kW option of the FAFNIR project followed by upgrade to 1.2MW beam power. The EU Roadmap milestone for 30dpa by end of 2026 is indicated by the solid lozenge whereas the present prediction for FAFNIR is end 2017.

Table 3 FAFNIR cost breakdown compared with IFMIF

|  | IFMIF US$M (2003) | FAFNIR US$M (2003) |
| --- | --- | --- |
| Project Management | 75.6 | 23.2 |
| Test facilities | 90.0 | 31.7 |
| Accelerator facitlities | 311.7 | 65.2 |
| Conventional facilities | 114.7 | 37.9 |
| Central C&I | 12.3 | 5.0 |
| Target facilities | 51.4 | 12.5 |
| Target development | - | 21.7 |
| **Total Capital (2003)** | **655.7** | **197.2** |
|  | IFMIF | FAFNIR |
| Total Capital (2012) | 818 | 246 |
| 20% Contingency 20% Manpower |  |  |
| **Total (2012 US$M)** | **1145** | **344** |

The comparison 2003 capital costs for FAFNIR and IFMIF are shown in Table 3 along with estimated equivalent 2012 construction values. (No allowance has been made for the accelerated inflation of raw materials in this estimate).The total commissioning costs of FAFNIR are estimated as US$41M and operational costs of US$15M per year assuming 70% availability, 24 hour operation in parallel with a target development program.

**4 Conclusion**

The case for constructing a reduced intensity, reduced risk 14MeV neutron source has been argued from an assessment of the regulatory requirements assuming defense in depth strategy and suitable formulation of the safety case. This removes the necessity to test the most highly irradiated in-vessel components to a full lifetime exposure, so that the purpose of the neutron source becomes proving investment protection rather than regulatory compliance. It is shown that the priority for such a source is to:

(i) identify new phenomena associated with 14MeV neutron irradiation

(ii) provide significant contributions to the population of the engineering materials database and eliminate unsuitable candidate materials

(iii) validate and calibrate fission and ion irradiation techniques and advance the materials modeling capability for fusion without compromising the validity of the neutron spectrum.

The facility would be of lower intensity than IFMIF but also of lower technological risk, exploiting existing or near term technologies to avoid lengthy R&D programs and challenging specifications. It is not the intention that this facility replace IFMIF but rather it will provide an intermediate step between IFMIF and existing irradiation facilities, which are generally not suited to fusion-relevant applications, and will generate data to support an extensive program of modeling to advance the understanding of irradiation effects on materials in a fusion environment and thus enable early elimination of unsuitable candidates.

To meet the requirements of the EU Fusion Roadmap of 30dpa irradiation by the end of 2026 the program will need to be aggressive and construction will need to begin in a timely manner. Estimates for the cost indicate approximately 30% of those of IFMIF. Such a facility would begin the population of the materials engineering database and would allow early elimination of unsuitable material candidates, enabling a more focused and cost effective use of the subsequent IFMIF facility.

**ACKNOWLEDGMENT**

This work was part-funded by the RCUK Energy Programme under grant EP/I501045. To obtain further information on the data and models underlying this paper please contact PublicationsManager@ccfe.ac.uk